Short Paper*
# ASIAVR: Asian Studies Virtual Reality Game a Learning Tool

Kenn Migan Vincent C. Gumonan
Information Technology Department, Northern Bukidnon Community College
kenngumonan@gmail.com
(corresponding author)

Aleta C. Fabregas
College of Computer and Information Sciences, Polytechnic University of the Philippines
aletfabregas@gmail.com

## Abstract

*Purpose* – The study aims to develop an application that will serve as an alternative learning tool for learning Asian Studies. The delivery of lessons into a virtual reality game depends on the pace of students. The developed application comprises several more features that enable users to get valuable information from an immersive environment.

*Method* – The researchers used Rapid Application Development (RAD) in developing the application. It follows phases such as requirement planning, user design, construction, and cutover. Two sets of questionnaires were developed, one for the teachers and another for the students. Then, testing and evaluation were conducted through purposive sampling to select the respondents.

*Results* – The application was overall rated as 3.56 which is verbally interpreted as very good. The result was based on the system evaluation using ISO 9126 in terms of functionality, usability, content, reliability, and performance.

*Conclusion* – The developed application meets the objectives to provide an alternative learning tool for learning Asian Studies. The application is well commended and accepted by the end-users to provide an interactive and immersive environment for students to learn at their own pace.

*Recommendations* – Further enhancement of the audio, gameplay, and graphics of the tool. Schools should take into consideration the adoption of the Asian Studies Virtual Reality as a good alternative tool for their teachers and students to teach and learn Asian Studies. The use of more 3D objects relevant to the given information to enhance game experience may be considered. A databank for the quiz questions that will be loaded into the game should also be considered.

*Research Implications* – The integration of modern technology in education has been a vital part of the learning process, especially when technological resources are available. Development and adaptation of this application will promote an alternative way of independent learning among students and will give them a better understanding of Asian Studies at their own pace.

*Keywords* – virtual reality, immersive environment, android-mobile game, learning tool, Asian studies

## INTRODUCTION

Technology and education are always in sync in terms of nurturing the students and attaining quality education; as technology keeps evolving, education on the other hand does the same. The methods and techniques in teaching the youth should always keep up with the latest trends and needs of today's generation. Computer games have already been around since the beginning of the computer era. The game industry continues to grow; it has become an industry that continues to shine through the years. The youth owns a large number of active and casual players of computer games on console and smartphones platform (Statista, 2018). Some are younger and even clever than adults in manipulating gadgets and playing games. Making computer games as an alternative way of learning would be beneficial especially for those students who are inclined in using technology as part of their everyday life.

The traditional teaching of lessons is always brought up to the classroom; it was proven effective since the early days. As the technology continues to evolve and gadgets like smartphones are getting cheaper each year, it paved the way for wider accessibility to the large population of the people in the world particularly the youth living in the Philippines. In 2017, there are more than 60 million smartphone users in the Philippines (Statista, 2018). This huge number of smartphone users has opened the way for e-learning accessibility which was given great attention and further development through the Virtual Reality (VR) technology, it has created a friendly human and machine interface. VR offers a variety of animated and interactive features which makes it an emerging essential tool in institutions (Abdelaziz, Riad, & Senousl, 2014). Virtual reality as a great tool used for instruction has become a trend in delivering the lesson and has helped diverse students to learn by scaffolding prior knowledge through the traditional use of images, this has provided an affordable way to support students through experiential and visual scaffolding (Pilgrim & Pilgrim, 2016).

An application that utilized low-cost and effective virtual reality technology for Grade 6 social studies was developed in the international school of Kazan, Republic of Tatarstan, Russia (Zantua, 2017). Though the study showed that the application produced positive results towards embedding lesson plans with virtual reality, the sample size of the respondents is just 20 and has not been assessed within the application. These gaps opened the need to create a virtual reality application for a higher grade level and a subject that is highly studied not only in the Philippines but in the entire world. Asian studies promote the rich culture, history, and people of Asia. ASIAVR was evaluated and highly commended by a larger number of respondents who appreciated the completeness of the application, not only on the lessons that can be found in the game but also the assessment part of the game which may also help the teachers in monitoring the progress of the students who opt to use the application as an alternative learning tool for learning Asian studies.

The study aims to develop an android application that will serve as an alternative learning tool for students in learning Asian studies through an interactive and immersive virtual reality environment, which can be played in an affordable smartphone and cheap VR headset. It can help technology inclined and visual learner students learn at their own pace. To attain this, the researchers interviewed the students and teachers about the issues and challenges they have encountered and gathered user requirements based on the lessons approved by the Department of Education to develop the application.

## LITERATURE REVIEW

The term "Virtual Reality" (VR) has been used for the first time in the '60s; VR has evolved in different manners becoming more and more similar to the real world. There are two different kinds of VR: non-immersive and immersive. VR in general showed how immersive VR in particular, has been used mostly for adult training in special situations or university students. It concludes outlining strategies that could be carried out to verify these ideas (Freina & Ott, 2015). It has captured the people's attention; a technology that has been applied in many sectors such as medicine, industry, education, video games, or tourism. Its largest field of scope has been leisure and entertainment. Over the years VR cost is declining, giving an affordable quality product that is available in the market. (Neff 2015). Other recent technological innovations, including the rapid adoption of smartphones by the society, have facilitated the access to virtual reality and augmented reality of anyone. Big companies like Apple, Facebook, Magic Leap, and Samsung, have increased their investment to make these technologies to improve their accessibility in the coming years. The huge possibilities of accessible virtual technologies will make it possible to break the boundaries of formal education (Gutiérrez et al., 2017).

The medium for creating a customized virtual reality world could range from creating a video game to having a virtual stroll around the universe, from walking through in a dream house to traverse on an alien planet (Mandal, 2013). Utilization of the Unity 3D software, a multi-platform game engine for creating games particularly Virtual Reality that provides an interactive walkthrough where users can experience and freely explore the virtual environment (Ling, 2015); a concept that can free humans from the restrictions of time and space by allowing humans to exist virtually to remote locations without travel is called Human Augmentation (AH). It augments humans not only in their senses and intellect but also in their abilities and motions to rise above time and space, which is termed as telexistence (Tachi, 2013).

An example of a virtual reality application used in education is an application for learning astronomy named Xolius, it was presented and evaluated by 20 students and 5 educators through interviews. It was found that VR is especially effective in subjects where an interactive environment is needed. VR also offers an immersive experience, involvement, and promoting active learning in comparison to the mobile application (Hussein & Natterdal, 2015). Another application that focused on artifact-models that are virtually endowed with different properties and metadata helped the users to learn things out using interacting with objects in the virtual reality environment (Neamtu & Comes, 2016). An application to promote situated and connected learning among nursing students Pharmacology Inter-Leaved Learning-Virtual Reality (PILL-VR) was developed as a simulation designed to enable practice and problem-solving in a safe environment (Dubovi, Levy, & Dagan, 2016).Lastly, embedding English as a Foreign Language (EFL) in the learning process of an oral language training into a generic 3D Cooperative Virtual Reality game shows that the designed virtual reality game trains the students' communication skills, evoking a high amount of speech and a qualitative linguistic output (Reitz, Sohny, & Lochmann, 2016).

Education needs to become more modular and move out of the classroom into informal settings like homes, and especially the internet. A certification based on these modules nationwide would permit technology to enter education more rapidly. Smaller nations may be more flexible in making these very disruptive changes (Psotka, 2013). Virtual Reality is not the progress itself; it is a medium of success. It suggests that the impact of Virtual Reality is still limited, so it can be used as an addition to traditional classrooms and standard training wherein the learners are immersed in a virtual environment (Minocha, 2015). VR games support students' autonomous learning at their own pace and understanding provided with the task that they have to accomplish (Hu, Su, & He, 2016).

# METHODOLOGY

*Software Development*

Figure 1 depicts the Rapid Application Development (RAD) methodology that was used by the researchers to develop the android application (Rouse, 2016). It has four phases. In Phase 1 which is the requirements and planning, the researchers submitted a letter to conduct interview with the teachers and students of CITI Global College. Through the interview conducted, the need to develop an application as an alternative way of learning Asian Studies was found out. Discussion on the project requirements and deliberations on potential scope and issues were taken into consideration. The prototype of the application was developed in Phase 2 through the gathered information from the interview. To come up with the best version of the application, several iterations were made.

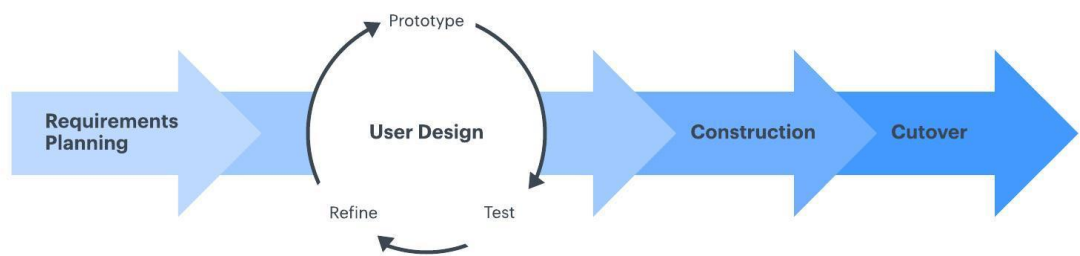

*Figure 1.* Rapid Application Development Methodology (Lucidchart, 2017)

In Phase 3 – Final Construction, the final integration of the application modules was conducted to finish the application. This included the lessons that were based on the Department of Education approved book for the Grade 7 students that was provided to the researchers by the Araling Panlipunan subject teachers. The researchers have also loaded the quiz questions into the game that was strictly evaluated by the teachers from the book provided. The researchers focused on conducting application testing and training using only an android smartphone as the only peripheral component, together with the low-cost headset to the end-users before the application was deployed in Phase 4 which is the cutover phase.

*System Testing and Evaluation*

The respondents of the study were the Grade 7 students and teachers of CITI Global College in Cabuyao, Laguna. They were chosen as respondents because the subject of Asian Studies falls under the Grade 7 curriculum as implemented by the Department of Education. There were (33) Grade 7 students and (2) social studies teachers for a total of (35) respondents. The researchers utilized the purposive sampling technique. It was used to select the subjects that best fit the study by own judgment and used when dealing with a limited target of subjects. The testing was conducted to each respondent individually using a smartphone with at least an android jelly bean operating system, that has a 4-7 inches screen size which is suitable for the generic virtual reality headset.

The tool used by the researchers in gathering data was the questionnaire and system evaluation using ISO 9126 in terms of functionality, usability, content, reliability, and performance. Table 1 shows the components of the evaluation questionnaire that were validated by the panel of oral examiners during the proposal stage and before the conduct of testing.

Table 1. Components of the evaluation questionnaire

| Criteria | Indicators |
| --- | --- |
| Functionality | Functionality |
|  | Provision of comfort and convenience |
|  | User-friendliness |
| Usability | Usability |
|  | Ease of learning |
|  | Efficiency of use |
|  | Memorability |
|  | Error frequency and severity |
|  | Subjective satisfaction |
| Content | Content |
|  | Updatability |
|  | Presentation |
| Reliability | Reliability |
|  | Absence of failures |
|  | Accuracy in performance |
| Performance | Trial 1 |
|  | Trial 2 |
|  | Trial 3 |

To measure the users' level of acceptance perspective and software evaluation items, a 5-point scale was used. The average weighted mean was computed and interpreted as (1) 4.20-5.00: Excellent; (2) 3.40-4.19: Very Good; (3) 2.60-3.39: Good; (4) 1.80-2.59: Fair; and (5) 1.00-1.79: Poor. The results are presented in the succeeding sections of this paper.

# THE PROPOSED SYSTEM

*System Architecture*

Figure 2 depicts the system architecture of the developed game that was compiled into an android application through the Unity 3D game engine application wrapper. It can be installed in a smartphone with a 4 to 7-inch screen size, and at least an Android jelly bean or higher Operating System. The game needs to be initialized before putting it inside the VR headset that may enable the student to experience and learn Asian Studies through Virtual Reality. Upon installation in the Android device, each player can only have one account in the game. The player needs to select a lesson; by default, only lesson 1 is loaded, if the game will be played for the first time. Succeeding lessons can be unlocked upon completion of the current lesson; in the game scene, the player has to move around the stage to gather objects like keys, flags, and other relevant game objects to complete the task. Once the player collides with the game object, the information will appear. After completing the stage, the player will be directed in the quiz stage wherein the player needs to choose the letter of choice to answer the question.

The player can also choose to play the previous lessons. Game progress will be saved directly to the built-in database of the game. The player name, quiz score per stage, and time duration of playing will be submitted after each quiz given that the device is connected to the internet. The teacher can view the submitted scores of the students through the website.

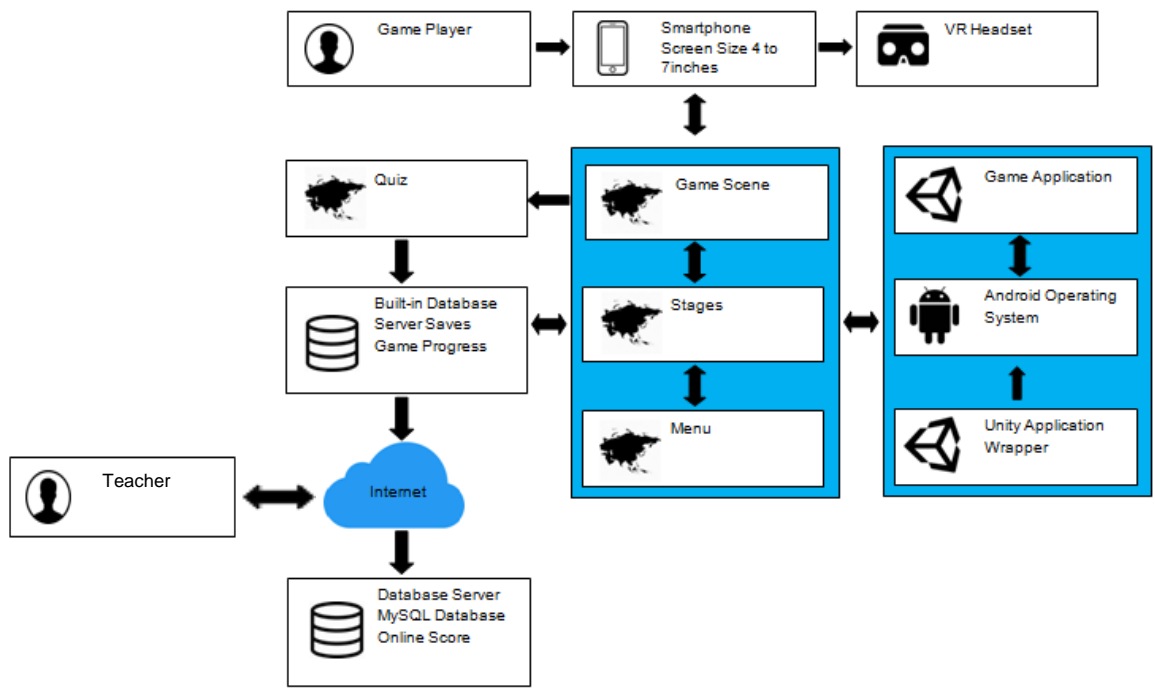

*Figure 2.* ASIAVR System's Architecture

*User Activities*

Figure 3 shows the scene where the user will access the game; it shows the main concept of the game. The player has to click the play button to proceed. Figure 4 shows the scene where the user will need to enter his/her last name and first name as player information that will be needed later on when the quiz score will be submitted in the online database.

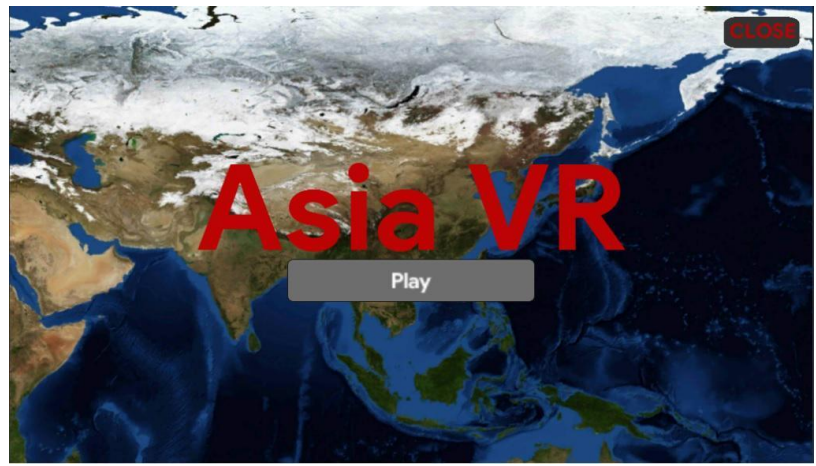

*Figure 3.* Title Menu Scene

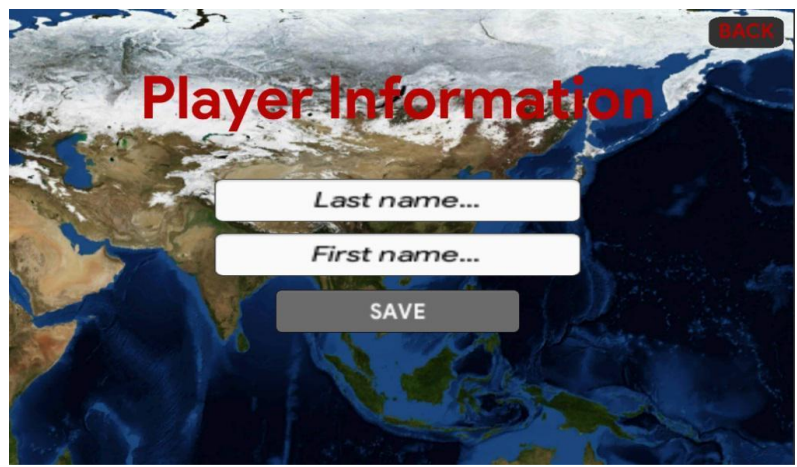

*Figure 4.* Player Information Scene

There are 8 stages as shown in Figure 5; the player's name will display on top. The player needs to select a particular stage and the game scene will be loaded right after. Playing the game for the first time will only allow the player to play the first stage which is Asia and all the other stages are locked, upon completion of each stage a new lesson will be unlocked. Upon selection of a particular stage in the game, a loading scene will appear to give an ample time of 10 seconds for the player to place the smartphone into the VR headset.

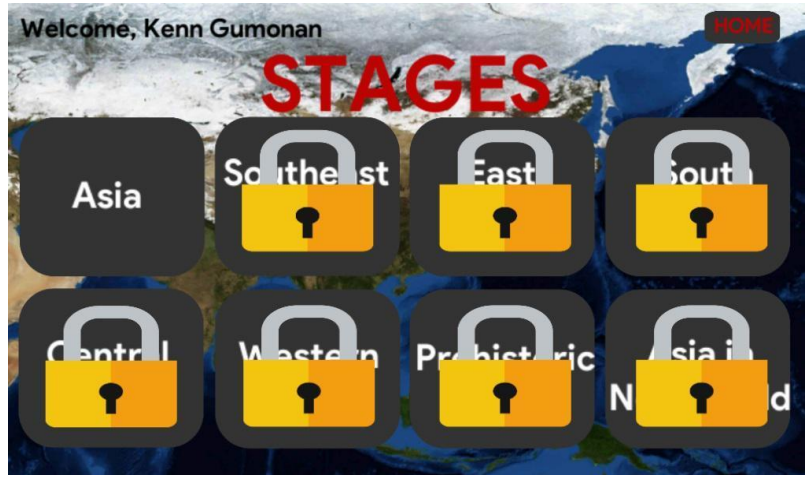

*Figure 5.* Stages Scene

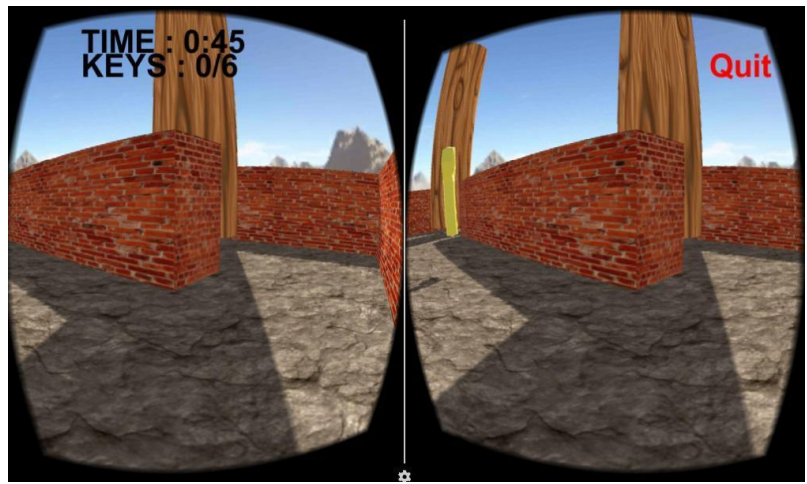

*Figure 6.* Game Scene

In Figure 6, the stage selected will load after the time runs out in the loading scene and the game time will start. The VR game environment complements the given lesson in Asian Studies, the player will have to navigate throughout the scene and look for objects to collect such as keys, country flags, and other game objects to finish the stage task; there is a game timer that will monitor the time duration of the player in minutes and seconds. The player may also choose to quit the game at any given point in time. Figure 7 shows the information scene that will load once the player bumps into a map or other relevant game objects while virtually touring around the game. Useful information relevant to the given lesson will be displayed on the screen. Figure 8 shows the quiz scene that will load after completing the task in each stage; each stage has a certain number of items for the quiz. All of the questions in the game came from the book approved by the Department of Education and was validated by the subject teachers in the school where the study was conducted. The player needs to collide the letter choice to answer the question.

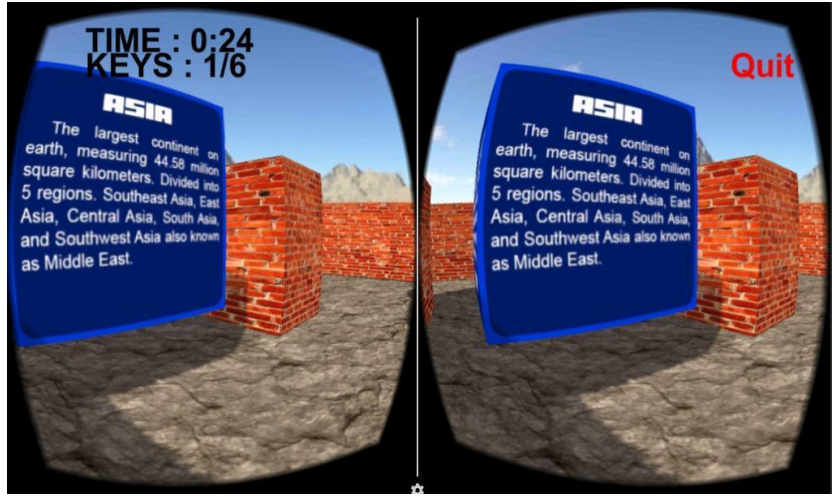

*Figure 7.* Information Scene

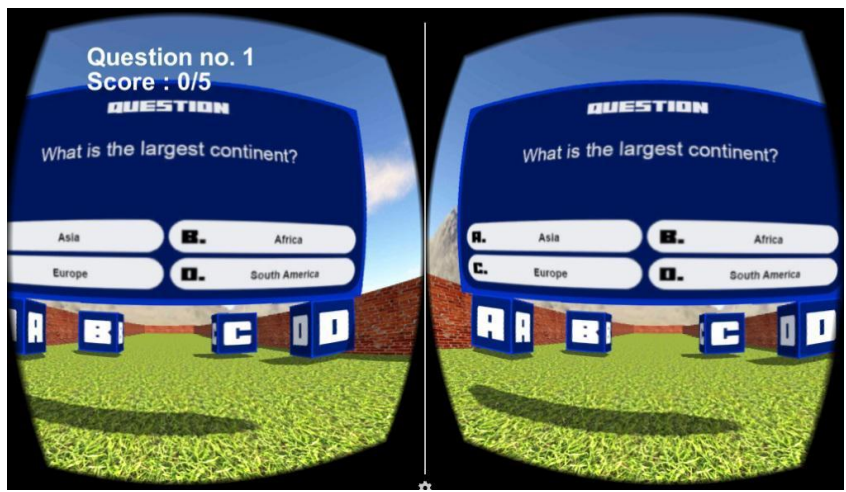

*Figure 8.* Quiz Scene

Figure 9 shows the submit score scene that will appear once the student is done answering all the quiz questions in each stage; the mobile device needs to have an internet connection for the player to submit the score upon clicking the submit button.

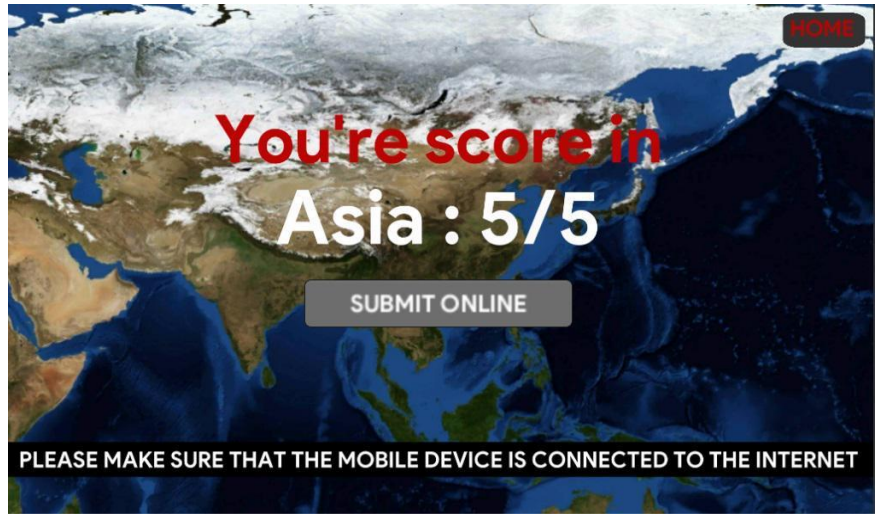

*Figure 9.* Submit Score Scene

*Figure 10.* Online Scores Scene

Figure 10 shows the submitted scores of the students that will be saved in the online database server of the ASIAVR wherein the name of the student, stage, score, time duration, and date played will be recorded online for the teachers to easily check and monitor the student progress.

**RESULTS AND DISCUSSIONS**

Table 2 shows the level of acceptance of the ASIAVR: Asian Studies Virtual Reality Game a Learning Tool. As observed from the table, the functionality of the application has 3.58 evaluated mean, which is interpreted as *very good*. This means that the application provides complete functionality that is accurate and secured. The usability of the application was also evaluated with a mean of 3.57 which is also interpreted as *very good*. This means that the application is easy to use and navigate, and it has an intuitive design. The content was graded with a mean of 3.61 which is interpreted as *very good*. It denotes that the information found in the game is accurate. The reliability of the application has a 3.44 evaluated mean, which is interpreted as *very good*. This means that the application can recover during unprecedented program errors, and can withstand failures when used. Performance has a mean of 3.61 with a verbal interpretation of *very good*. This means that the application has a very good presentation before the end-users. The overall average of the different aspects is 3.56 which is verbally interpreted as *very good*. which suits the criteria for an application to provide an alternative way of learning Asian Studies.

Table 2. Overall Acceptance of the Developed Application

| Criteria | Mean | Verbal Interpretation |
|---|---|---|
| Functionality | 3.58 | Very Good |
| Usability | 3.57 | Very Good |
| Content | 3.61 | Very Good |
| Reliability | 3.44 | Very Good |
| Performance | 3.61 | Very Good |
| **AWM** | **3.56** | **Very Good** |

## CONCLUSIONS AND RECOMMENDATIONS

Based on the preceding findings of the study, the respondents agreed that the Asian Studies Virtual Reality Game as a Learning tool was functional and could help students and teachers in teaching and learning Asian Studies. The overall performance of the game was very good in terms of the level of acceptance in functionality, usability, content, reliability, and performance.

For the future development of the study, it is recommended that there should be an improvement in the audio, gameplay, and graphics of the tool. Schools should take into consideration the adoption of the Asian Studies Virtual Reality game as a good alternative tool for their teachers and students to teach and learn Asian Studies. The use of more 3D objects relevant to the given information to enhance game experience may be considered. Lastly, a databank for the quiz questions that will be loaded into the game should also be considered.

# IMPLICATIONS

The availability of technology has opened up a new way for students to learn. The integration of modern technology in education has been a vital part of the learning process. Development and adaptation of this application will promote an alternative way of independent learning among students and will give them a better understanding of Asian Studies at their own pace.

# ACKNOWLEDGEMENT

The researchers would like to thank our almighty God for the abundant grace and gift of knowledge, their family, friends, and colleagues for their never-ending love and support. The authors are forever grateful to the owner of CITI Global College, Dr. Edwin C. Buraga for his utmost support; to the junior high school principal Mr. John Irwin E. Delmo for his approval to conduct the study, the researchers admired the gesture and open-mindedness of the school administrators in the application of technology in the field of education; to the junior high school teachers Ms. Florabille I. Gumapon and Ms. Norileen Capuso, for the time and effort that they had given to the researchers in allowing them to conduct the study during their class hours and guiding them with the lessons and questions that needs to be applied in the game in accordance to the lessons that they teach to the students in Asian Studies as prescribed by the Department of Education; and to the Grade 7 students of CGC for their active participation in the game testing and generously answering the survey questions.

# REFERENCES

Abdelaziz, M., Riad, A. E., & Senousl, M. B. (2014). Challenges and issues in building virtual reality-based e-learning systems. *International Journal of e-Education, e-Management, and e-Learning,4*(4), 320-328. Retrieved from https://www.semanticscholar.org/paper/Challenges-and-Issues-in-Building-Virtual-System-Abdelaziz-El/1c771c3b1d08e6cd68776d9f7e5ab4ac4d7b7598

Dubovi, I., Levy, S., & Dagan, E. (2016). PILL-VR simulation environment for teaching medication administration to nursing students. In *11th Chais Conference for the Study of Innovation and Learning Technologies: Learning in the Technological Era*. Retrieved from https://www.researchgate.net/publication/301694042_PILL-VR_Simulation_Learning_Environment_for_Teaching_Medication_Administration_to_Nursing_Students/link/5723134208ae586b21d56807/download

Freina, L., & Ott, M. (2015). A literature on immersive virtual reality in education: state of the art and perspectives. *eLearning and Software for Education, 1, 133*. Retrieved from http://citeseerx.ist.psu.edu/viewdoc/summary?doi=10.1.1.725.5493

Gutiérrez, J., Mora, C., Diaz, B., & Marrero, A. (2017). Virtual technologies trends in education. *EURASIA Journal of Mathematics Science and Technology Education, 13*(2), 469-486. Retrieved from https://www.ejmste.com/download/virtual-technologies-trends-in-education-4674.pdf

Hu, X., Su, R., & He, L. (2016). The design and implementation of the 3d educational game on VR headsets. *International Symposium on Educational Technology (ISET)*, 53-56. Retrieved from https://ieeexplore.ieee.org/document/7685593

Hussein, M., & Natterdal, C. (2015). *The benefits of virtual reality in Education: A comparison study* [Bachelor Thesis, University of Gothenburg], University of Gothenburg Institutional Repository. Retrieved from https://gupea.ub.gu.se/bitstream/2077/39977/1/gupea_2077_39977_1.pdf

Ling, Y. M. (2015). *Virtual interactive interior walkthrough using UNITY3D* (unpublished manuscript). University Malaysia Sarawak Institutional Repository. Retrieved from https://ir.unimas.my/id/eprint/12231/1/Virtual%20interactive%20interior%20walkthrough%20using%20unity3d%20(24pgs).pdf

Lucidchart. (2017). *4 phases of rapid application development methodology*. Retrieved from https://www.lucidchart.com/blog/rapid-application-development-methodology#

Mandal, S. (2013). Brief introduction of virtual reality and its challenges. *International Journal of Scientific and Engineering Research*, 4(4), 304-309.

Minocha, S. (2015). The state of virtual reality in education –shape of things to come. *International Journal of Engineering Research,* 4(11), 596-598.

Neamtu, C., & Comes, R. (2016). Methodology to create digital and virtual 3d artefacts in archeology. *Journal of Ancient History and Archaeology,* 3(4), 65-74. Retrieved from http://jaha.org.ro/index.php/JAHA/article/download/206/170

Neff, J. (2015). *How virtual reality could change shopper marketing, b2b and more.* Retrieved from http://adage.com/article/digital/virtual-reality/299336/

Pilgrim, J. M. & Pilgrim, J. (2016). The use of virtual reality tools in the reading-language arts classroom. *The Journal of Literacy Education,* 4(2), 90-97. Retrieved from https://eric.ed.gov/?id=EJ1121641

Psotka, J. (2013). Educational technology and virtual reality as disruptive technologies. *Educational Technology & Society,*16(2), 69-80.

Reitz, L., Sohny, A., & Lochmann, G. (2016). VR-based gamification of communication training and oral examination in a second language. *International Journal of Game-Based Learning,*6(2)46-61.

Rouse, M. (2016). *Rapid Application Development (RAD)*. Retrieved from http://searchsoftwarequality.techtarget.com

Statista. (2018). *Number of smartphone users worldwide from 2014 to 2020*. Retrieved from https://www.statista.com/statistics/330695/number-of-smartphone-users worldwide/

Tachi, S. (2013). From 3D to VR and further to telexistence. In *2013 23rd International Conference on Artificial Reality and Telexistence (ICAT)*, *1,* 1-10.

Zantua, L. (2017). Utilization of virtual reality content in grade 6 social studies using affordable virtual reality technology. *Asia Pacific Journal of Multidisciplinary Research,* 5(2), 1-10. Retrieved from http://www.apjmr.com/wp-content/uploads/2017/05/APJMR-2017.5.2.2.01.pdf